\def\tld{truncated L\`evy distribution}
\def\iems#1{{\protect\index{#1|bold}}}
\def\aus#1{{\protect\index{aaauthor:#1}}}

\def\deltab{{\mbox{\boldmath$\delta$}}}

\def\aut#1{#1}
\def\ins#1{}
\def\iem#1{{\em #1}}

\def\comment#1{}

\def\cm#1{}

\def\>{\rangle}
\def\<{\langle}
\def\comment#1{}
\def\ind#1{#1}

\documentstyle[pra,epsf,aps]{revtex}
\begin{document}

\title{
Option Pricing
from Path Integral for Non-Gaussian Fluctuations.\\
Natural Martingale and Application to
Truncated L\`evy Distributions%
\thanks{
kleinert@physik.fu-berlin.de,
{}~ http://www.physik.fu-berlin.de/\~{}kleinert }}
\author{
Hagen Kleinert\\[1mm]
Institute for Theoretical Physics, FU-Berlin,
Arnimallee 14, D-14195, Berlin, Germany
}
\maketitle
\begin{abstract}
Within a path integral formalism
for
 non-Gaussian
price fluctuations
we set up a simple stochastic calculus
  and {\em derive\/}
a natural martingale
for option pricing
from the wealth balance of options, stocks, and bonds.
The resulting formula is evaluated
for \tld{}s.
\end{abstract}

\section{Introduction}

Financial markets exhibit non-Gaussian price
fluctuations. For this reason they are
incomplete markets
and mathematicians are unable
determine a unique riskless martingale distribution
to calculate fair option prices.

As pointed out in  Pareto's
 19th century work \cite{PAR} and reemphasized by
Mandelbrot
 in the 1960s,
 the logarithms of assets prices
in
financial markets
do not fluctuate with
Gaussian
distributions,
but  possess much larger  tails
which may be approximated by  various other
distributions such as
truncated L\'evy distributions \cite{MB1,MES,KOP,BP},
Meixner distributions \cite{Gri,Mei},
 generalized hyperbolic distributions
or simplifies versions thereof
\cite{N1,N2,N2a,N3,N4,N5,N6,N7,N8,N9,N9b,N10,N11,N12,N13,N14,N15,N16,N17,N18,N19,N20,N21,N22,N23}.
This
has the unpleasant consequence
that the
associated stochastic differential equations
 cannot be treated with
the popular Ito calculus. The mathematicians call such markets
{\em incomplete\/}, implying that
there are different
choices of riskless martingale distributions with which one
can calculate option prices.
Many of these
have been discussed in the literature, and
mathematicians have invented various sophisticated
criteria under which
one would be preferable over the others
for calculating financial risks.
Davis, for instance,  has introduced a so-called \iem{utility function}
\cite{Davis} which is supposed to select optimal martingales
for different purposes.
\comment{There exists also a so-called minimal martingale
and the precise virtues of this choice
in these discussions is hard to
understand for the non-mathematician.
}

In this paper we want to point out that in a path integral formulation
of the problem,
a straightforward extension
of the old chain of arguments
which led Black-Scholes
to their famous formula
produces  a specific
simple martingale
which is different from presently popular versions
based on Esscher transforms.
Before we come to this we
briefly review the
derivation of a
stochastic calculus for non-Gaussian processes \cite{NITO}
from path integrals which extends
Ito's calculus in the simplest way.

\section{Gaussian Approximation to
Fluctuation Properties of Stock Prices}
Let $S(t)$ denote the price of some stock.
On the average, stocks grow exponentially,
and a first
approximation
considers
a stock price as
an exponential $S(t)=e^{x_S(t)}$
of a fluctuating variable
$x(t)$ which obeys a stochastic differential equation
\begin{equation}
\dot x_S(t)=r_{x_S}+\sigma \eta (t),
\label{@stochdequstox}\label{@stdex}\end{equation}
where
$ \eta (t)$ is a white noise of unit strength
with the correlation functions
\begin{equation}
\langle \eta(t)  \rangle =0,~~~~~~
\langle \eta(t) \eta(t') \rangle =
 \delta (t-t').
  \label{stocknoise}
\end{equation}

 The
stock price
$
 S(t)
$
 itself satisfies the stochastic differential equation
for exponential growth
\begin{equation}
\frac{\dot S(t)}{S(t)}=r_S+ \sigma  \eta (t),
\label{@stochdequsto}\end{equation}
where
The standard deviation $ \sigma$ is a measure for
the volatility of the stock price, which
defined by the
 expectation value
\begin{eqnarray}
 \Bigg \langle\left[ \frac{\dot S(t)}{S(t)}\right]^2   \Bigg \rangle
= \sigma ^2            .
\label{@volati}\end{eqnarray}
The  rate constants in Eqs.~(\ref{@stdex})
and (\ref{@stochdequsto})
differ from one another
due the stochastic nature of  $x_S(t)$ and $S(t)$.
According to  Ito's rule,
  we may expand
\begin{eqnarray}
\dot x_S(t)&=&
\frac{dx_S}{dS}\dot S(t) +\frac{1}{2}\frac{d^2x_S}{dS^2}\dot S^2(t)\, dt+\dots\nonumber \\
&=&\frac{\dot S(t)}{S(t)}-\frac{1}{2}\left[ \frac{\dot S(t)}{S(t)}\right]^2  dt+\dots
,
\label{@Itonew0}\end{eqnarray}
and replace the last term by its expectation
value (\ref{@volati}):
\begin{eqnarray}
\dot x_S(t)&=&
\frac{\dot S(t)}{S(t)}-\frac{1}{2}
 \Bigg \langle\left[ \frac{\dot S(t)}{S(t)}\right]^2   \Bigg \rangle dt\nonumber \\
&=&\frac{\dot S(t)}{S(t)}-\frac{1}{2} \sigma^2+\dots~.
\label{@Itosrule}\end{eqnarray}
Inserting  Eq.~(\ref{@stochdequsto}),
we obtain the well-known Ito relation
\begin{eqnarray}
r_ {x_S}=r_S- \frac{1}{2}  \sigma ^2.
\label{@itoreln}\end{eqnarray}
In praxis, this relation implies
that if we fit
a straight line
 through a plot of the logarithms
of stock prices, the forward
extrapolation of the average stock price is
given by
\begin{equation}
S(t)=S(0)\,e^{r_St}=
S(0)\,e^{(r_{x_S}+ \sigma ^2/2)t}.
\label{@forwS}\end{equation}

\section{Non-Gaussian Distributions}
The description
of the fluctuations
of the logarithms of the stock prices
around the linear trend
by a Gaussian distribution
is  only a rough
approximation to the real
stock prices.
Given a certain time scale, for instance
days,
they follow
some distribution
$\tilde D(x)$
which is the Fourier transform
of some function
$D(p)$, which we shall write as an exponential
$e^{-H(p)}$:
\begin{equation}
\tilde D(x)=
\int \frac{dp}{2\pi}e^{ipx}D(p)=
\int \frac{dp}{2\pi}e^{ipx-H(p)}.
\label{@}\end{equation}
In general, $H(p)$ will have a power series expansion
\begin{equation}
 H(p)=
ic_1p+\frac{1}{2} c_2\, p^2
-i\frac{1}{3!}c_3p^3-\frac{1}{4!}c_4\,p^4
+i\frac{1}{5!}c_5p^5
+\dots,
\label{@cumuatsHa}\end{equation}
The coefficients $c_n$ in the expansion (\ref{@cumuatsHa}) are the {\em cumulants\/}
of the distribution $\tilde D(x)$,
from which they can be obtained by the
 connected expectation values
\begin{equation}
c_1=\langle x\rangle_c \equiv
\langle x\rangle \equiv \int dx\,x\,\tilde D(x),  ~~~
c_2=\langle x^2\rangle_c\equiv
\langle x^2\rangle-
\langle x\rangle^2
 \equiv
 \int dx\,(x-\langle x\rangle )^2\equiv  \sigma ^2\,\tilde D(x),\dots~.
\label{@}\end{equation}

By analogy with mechanics, we call $H(p)$ the {\em Hamiltonian\/} of the
fluctuations.
A Gaussian Hamiltonian
$H(p)= \sigma ^2p^2/2$
coincides with the mechanical energy of a
free particle of mass $1/ \sigma ^2$.
Non-Gaussian Hamiltonians
are standard in elementary particle physics.
A relativistic particle, for example, has a Hamiltonian $H(p)= \sqrt{p^2+m^2}$,
and there is no problem in defining and solving the associated path integral
\cite{PI}.

We also introduce a modified Hamiltonian
$\bar H(p)$, whose Fourier transform has zero average.
Its expansion (\ref{@cumuatsHa})
has the linear term $ic_1p$ subtracted.  We also
define
a  modified Hamiltonian $H_r(p)$
by adding to $\bar H(p)$
a linear term $irp$, i.e.,
\begin{equation}
\bar H(p)\equiv H(p)-H'(0)p,~~~
 H_r(p)\equiv
\bar H(p)+irp=
 H(p)-H'(0)p+irp.
\label{@defhbar}
\label{@cumuatsHar}
\end{equation}

\subsection{Path Integral for Non-Gaussian
Fluctuations}

It is easy to calculate the properties of a
process (\ref{@stdex}) whose fluctuations
are distributed according to a
general non-Gaussian distribution.
If we assume the rate $r_{x_S}$
to coincide with the
linear coefficient
$c_1$ in the expansion (\ref{@cumuatsHa}),
such that we can abbreviate
$H_{r_{x_S}}(p)$
by
 $H(p)$,
the stochastic differential equation
reads
\begin{equation}
\dot x_S(t)=\eta (t),
\label{@stdeqet}\end{equation}
and the probability distribution of the endpoints
of paths starting at a certain initial
point   is given by the path integral
\begin{eqnarray}
P(x_b t_b|x_at_a)=
\int {{\cal D}  \eta }
\int {{\cal D}  x}
\exp\left[ -\int _{t_a}^{t_b}dt \, \tilde H( \eta (t))\right]
 \delta [\dot x- \eta ],
\label{@18spiLP}
\end{eqnarray}
with the initial condition
$x_S(t_a)=x_a$. The final point is, of course, $x_b=x_S(t_b)$.
The function $\tilde H( \eta )$ is defined by
the negative logarithm of  the non-Gaussian distributions $\tilde D (x)$, such that
\begin{equation}
e^{-\tilde H(x)}\equiv  \tilde D(x).
\label{@}\end{equation}
The
path integral is defined \'a la Feynman \cite{PI}
by slicing the time axis at times $t_n= \epsilon n$ with $n=0,\dots,N$,
and integrating over all $x(t_n)$. At the end, the limit $N$ is taken.
In this way, we select from the space of all
fluctuating paths
a well-behaved set  of measure zero, which is in most cases sufficient to
obtain the correct limit $N\rightarrow \infty$, in particular it will be for the typical non-Gaussian distributions encountered in stock
markets. This is completely analogous
to Riemann's  procedure of defining ordinary integrals,
which are approximated by
sums over values of a function on a set of measure zero.

With the help of
the path integral
(\ref{@18spiLP}),
the  correlation functions of the noise in the path integral
(\ref{@18spiLP})
can be found
 by  straightforward functional differentiation.
For this purpose,
 we
 express the noise distribution
 $P[ \eta ]\equiv \exp\left[ {-\int_{t_a}^{t_b} dt\,\tilde H(\eta(t) )}\right] $
as a Fourier path integral
\begin{eqnarray}\!\!\!\!\!\!\!\!\!
P[ \eta]= \int {{\cal D}  \eta } \!
\int \frac{{\cal D}  p }{2\pi}
\exp\left\{
\int _{t_a}^{t_b}\!dt\left[ip(t) \eta (t)-  H( p(t))\right] \right\},
\label{@18spiLP1n}
\end{eqnarray}
and note that the correlation functions
can be obtained from
the functional derivatives
\begin{eqnarray}
&&\!\!\!\!\!\!\!\!\!\!
\langle  \eta (t_1)\cdots \eta (t_n)\rangle =(-i)^n
\int {{\cal D}  \eta } \!
\int \frac{{\cal D}  p }{2\pi}
\left[ \frac{ \delta }{ \delta p(t_1)}                    \cdots
\frac{ \delta }{ \delta p(t_n)}
e^{i \int _{t_a}^{t_b}\!dt\,p(t) \eta (t)} \right]
e^{- \int _{t_a}^{t_b}\!dt\,  H( p(t)) }.   \nonumber \\&&
\label{@18spiLP1n1}
\label{@}\end{eqnarray}
After $n$  partial integrations, this becomes
\begin{eqnarray}
\!\!\!\!\!\!\!\!\!\!\!\!
\langle  \eta (t_1)\cdots \eta (t_n)\rangle &=&
i^n
\int {{\cal D}  \eta } \!
\int \frac{{\cal D}  p }{2\pi}  e^{i \int _{t_a}^{t_b}\!dt\,p(t) \eta (t)}
\, \frac{ \delta }{ \delta p(t_1)}                    \cdots
\frac{ \delta }{ \delta p(t_n)}
 e^{- \int _{t_a}^{t_b}\!dt\,  H( p(t)) }
       ~~~~~~\nonumber \\[2mm]
&=&i^n
\left[ \frac{ \delta }{ \delta p(t_1)}                    \cdots
\frac{ \delta }{ \delta p(t_n)}
 e^{- \int _{t_a}^{t_b}\!dt\,  H( p(t)) }
    \right]_{p(t)\equiv 0} . ~~~~~~~~~
\label{@18spiLP1n1}
\end{eqnarray}
By expanding the exponential  $
 e^{- \int _{t_a}^{t_b}\!dt\,  H( p(t)) }$ in a power series using
 (\ref{@cumuatsHa}),
it is straightforward to calculate
\begin{eqnarray}
\langle  \eta (t_1)  \rangle
&\equiv&     Z^{-1}
\int {{\cal D}  \eta } \,\eta (t_1)
\exp\left[ -\int _{t_a}^{t_b}dt \, \tilde H( \eta (t))\right]
=c_1,\label{@correlfu1} \\
\langle  \eta (t_1) \eta (t_2) \rangle
&\equiv&     Z^{-1}
\int {{\cal D}  \eta } \,\eta (t_1) \eta (t_2)
\exp\left[ -\int _{t_a}^{t_b}dt \, \tilde H( \eta (t))\right]
\nonumber \\
&=&c_2 \delta (t_1-t_2)+c_1^2,
\label{@correlfu2}\\
\langle
 \eta (t_1) \eta (t_2) \eta (t_3)
 \rangle
&\equiv&  Z^{-1}
\int {{\cal D}  \eta }\,
 \eta (t_1) \eta (t_2) \eta (t_3)
\exp\left[ -\int _{t_a}^{t_b}dt \, \tilde H( \eta (t))\right]
\nonumber \\
&=&
c_3
\delta (t_1-t_2)
\delta (t_1-t_3)
  \\
&+&c_2c_1
\big[
\delta (t_1-t_2)+\delta (t_2-t_3)+
\delta (t_1-t_3)
\big] ,
\nonumber \\
\langle
 \eta (t_1) \eta (t_2) \eta (t_3) \eta (t_4)
 \rangle
&\equiv&  Z^{-1}
\int {{\cal D}  \eta }\,
 \eta (t_1) \eta (t_2) \eta (t_3) \eta (t_4)
\exp\left[ -\int _{t_a}^{t_b}dt \, \tilde H( \eta (t))\right]
\nonumber \\&=&
c_4
\delta (t_1-t_2)
\delta (t_1-t_3)
\delta (t_1-t_4)
\nonumber \\&+&
c_3c_1
\big[
\delta (t_1-t_2)\delta (t_1-t_3)
+3\,{\rm cyclic~perms}
\big]
 \label{@correlfu} \\
&+&c_2c_1^2
\big[
\delta (t_1-t_2)+5\,{\rm pair~terms}\big]
\nonumber \\
&+&
Ac_2^2
\big[
\delta (t_1-t_2)\delta (t_3-t_4)
\delta (t_1-t_3)\delta (t_2-t_4)
\delta (t_1-t_4)\delta (t_2-t_3)\big] , \nonumber
\end{eqnarray}
where   $Z$ is the normalization integral
\begin{eqnarray}
Z\equiv
\int {{\cal D}  \eta } \,\exp\left[ -\int _{t_a}^{t_b}dt \, \tilde H( \eta (t))\right].
 \label{@}\end{eqnarray}
The higher
 correlation functions are obvious generalizations
of these equations.
The different contributions
on the right-hand side of
 Eqs.~(\ref{@correlfu1})--(\ref{@correlfu})
are distinguishable
by their connectedness structure.

\begin{figure}[htbp]
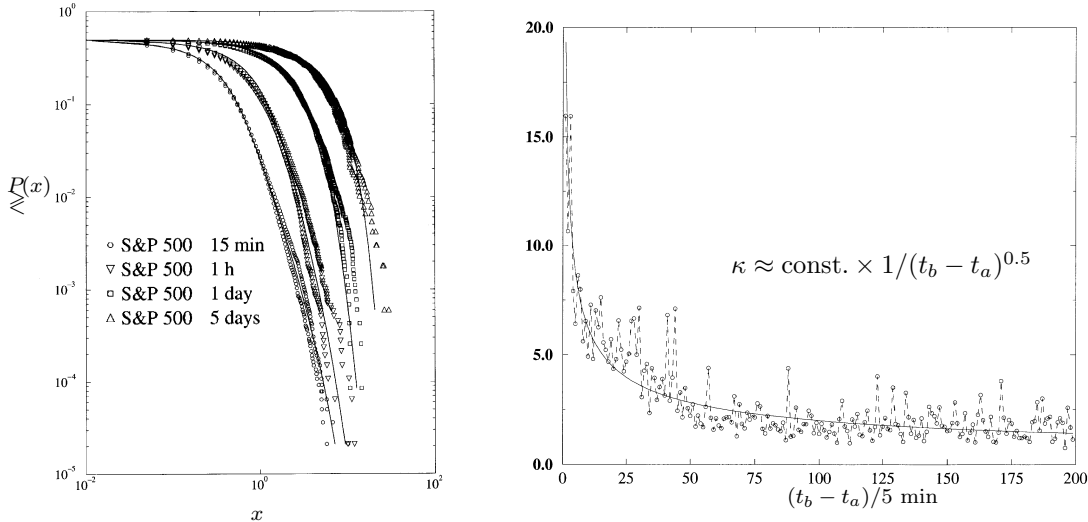

\vspace{-.4cm}
\unitlength .36mm
\input convsp.tps \\[10mm]
\caption[Left-hand side: Cumulative distributions obtained
from repeated convolutions
of the 15-min distribution.
Right-hand side:
Falloff of kurtosis is slower than the expected  $1/(t_b-t_a)$
from
convolution property]{Left-hand side: Cumulative distributions obtained
by repeated convolutions
of the 15-min distribution (from Ref.~\cite{BP}).
 Apart from the tails, the
semigroup property (\ref{@convprtop})
is reasonably well satisfied. Right-hand side:
Falloff of kurtosis is slower than $1/(t_b-t_a)$
 expected
from
convolution property (see Ref.~\cite{COND}). }
\label{@convsp}\end{figure}

An important property of the probability
(\ref{@18spiLP}) is that it satisfies
the semigroup property  of
 path integrals
\begin{eqnarray}
P(x_c t_c|x_at_a)=
\int dx_b\,P(x_c t_c|x_bt_b)P(x_b t_b|x_at_a).
\label{@convprtop}\end{eqnarray}
The the experimental
asset distributions do  satisfy this property approximately.
For truncated L\'evy distributions
this is shown in Fig.~\ref{@convsp}.
 Apart from the far ends of the tails, the
semigroup property (\ref{@convprtop})
is reasonably well satisfied.

\subsection{L\'evy-Khintchine Formula}
If is sometimes
useful to represent
 the Hamiltonian
in the form of a Fourier integral
\begin{equation}
H(p)=-\int dx\, e^{ipx}F(x).
\label{@LevyKhich}\end{equation}
Due to the the special significance
of the linear  term in $H(p)$
governing the drift,
this
is usually subtracted out of the integral
by rewriting (\ref{@LevyKhich}) as
\begin{equation}
H_r(p)=irp+\int dx \left(e^{ipx}-1-ipx
\right)F(x).
\label{@}\end{equation}
 The first subtraction ensures the property $H_r(0)=0$
which guarantees the unit normalization of the distribution.
This subtracted representation is known as the
\iem{L\'evy-Khintchine formula},
and the function $F(x)$ is the so-called
\iems{L\'evy measure}
of the distribution.
Some people also subtract out the quadratic term
and write
\begin{equation}
H_r(p)=irp+ \frac{ \sigma ^2}2p^2+\int dx \left(e^{ipx}-1
\right)\bar F(x).
\label{@LKSUB}\end{equation}
and use a function $\bar F(x)$ which has no
first and second moment (
i.e.,
$\int F(x)x=0,\,
\int F(x)x^2=0$.
to avoid redundancy in the representation.

  \subsection{Fokker-Planck-Type Equation}
\label{@FPLEQ}
The $ \delta $-functional
may be represented
by a Fourier integral leading to
\begin{eqnarray}
P(x_b t_b|x_at_a)=
\int {{\cal D}  \eta }
\int \frac{{\cal D}  p }{2\pi}
\exp\left\{ \int _{t_a}^{t_b}dt\left[i p(t)\dot x_S(t)-ip(t) \eta (t)- \tilde H(  \eta  (t))\right] \right\}
\label{@18spiLP1}       .
\end{eqnarray}
Integrating out the noise variable
$ \eta (t)$ at each time $t$
 we obtain
\begin{eqnarray}
P(x_b t_b|x_at_a)=
\int \frac{{\cal D}  p }{2\pi}
\exp\left\{ \int _{t_a}^{t_b}dt\left[i p(t)\dot x_S(t)- H(  p  (t))\right] \right\}
\label{@18spiLP2}      .
\end{eqnarray}
Integrating over all $x_S(t)$ with  fixed end points
enforces a constant momentum
along the path, and we remain with a single
integral
\begin{eqnarray}
P(x_b t_b|x_at_a)=
\int \frac{{d}  p }{2\pi}
\exp\left[ -({t_b}-{t_a}) H(  p)+ip(x_b-x_a) \right]
\label{@18spiLP3}      .
\end{eqnarray}
From this representation
it is easy to verify that this probability
satisfies a Fokker-Planck-type equation
\begin{eqnarray}
\partial _tP(x_b t_b|x_at_a)=-H(-i\partial _x)
P(x_b t_b|x_at_a).
\label{@FPDEL}\end{eqnarray}
The general solution $\psi(x,t)$
of this differential equation
with the initial condition $\psi(x,0)$
is given by the path integral generalizing (\ref{@18spiLP})
\begin{eqnarray}
\psi(x,t)=
\int {{\cal D}  \eta }
\exp\left[ -\int _{t_a}^{t_b}dt \, \tilde H( \eta (t))\right]
 \psi\left( x-\int_{t_a}^t dt' \eta(t')\right).
\label{@18spiLPg}
\end{eqnarray}
To verify  that this
satisfies indeed
the Fokker-Planck-type equation  (\ref{@FPDEL})
we consider $\psi(x,t)$ at a slightly later time
$t+ \epsilon $
and expand
\begin{eqnarray}~~~\!\!\!\!\!\!\!\!\!\!\!\!\psi(x,t+ \epsilon ) ~~~~
&&
\!\!\!\!\!\!\!\!\!\!\!\!\!\!\!\!\!\!\!\!\!\!\!\!\!\!\!\!\!\!~~~~~
\hspace{9mm}
=\int {{\cal D}  \eta }
\exp\left[ -\int _{t_a}^{t_b}dt \, \tilde H( \eta (t))\right]
 \psi
\left( x-\int_{t_a}^t dt' \eta(t')-\int_{t}^{t+ \epsilon } dt' \eta(t')\right).
~~~~\nonumber \\
&&
\!\!\!\!\!\!\!\!\!\!\!\!\!\!\!\!\!\!\!\!\!\!\!\!\!\!\!\!\!\!~~~~~
\hspace{9mm}
=
\int {{\cal D}  \eta }
\exp\left[ -\int _{t_a}^{t_b}dt \, \tilde H( \eta (t))\right]  \left\{
 \psi
\left( x-\int_{t_a}^t dt' \eta(t')\right)
\right.\nonumber \\
&&\!\!\!\!\!\!\!\!\!\!\!\!\!\!\!\!\!\!\!\!\!\!\!\!\!\!\!\!\!\!~
\hspace{9mm}
~~~~
- \left.~~\psi '
\left( x-\int_{t_a}^t dt' \eta(t')\right)
\int_{t}^{t+ \epsilon } dt' \eta(t')
\right.\nonumber \\
&&\!\!\!\!\!\!\!\!\!\!\!\!\!\!\!\!\!\!\!\!\!\!\!\!\!\!\!\!\!\!~
\hspace{9mm}
~~~~+\left.\frac{1}{2} \psi ''
\left( x-\int_{t_a}^t dt' \eta(t')\right)
\int_{t}^{t+ \epsilon } dt_1 dt_2\,\eta(t_1)
\eta(t_2)
\label{@18spiLPgl1}
\right. \\
&&\!\!\!\!\!\!\!\!\!\!\!\!\!\!\!\!\!\!\!\!\!\!\!\!\!\!\!\!\!\!~
\hspace{9mm}
~~~~-\left.\frac{1}{3!} \psi '''
\left( x-\int_{t_a}^t dt' \eta(t')\right)
\int_{t}^{t+ \epsilon } dt_1 dt_2dt_3\,\eta(t_1)
 \eta(t_2)
 \eta(t_3)
\right.\nonumber \\
&&\!\!\!\!\!\!\!\!\!\!\!\!\!\!\!\!\!\!\!\!\!\!\!\!\!\!\!\!\!\!~
\hspace{9mm}
~~~~+\left.\frac{1}{4!} \psi ^{(4)}
\left( x-\int_{t_a}^t dt' \eta(t')\right)
\int_{t}^{t+ \epsilon } dt_1 dt_2dt_3dt_4\,\eta(t_1)
 \eta(t_2)
 \eta(t_3)
 \eta(t_4)   +\dots~
\right\}.\nonumber
\end{eqnarray}
Using the
correlation functions
(\ref{@correlfu1})--(\ref{@correlfu})
we obtain
\begin{eqnarray}
\psi(x,t+ \epsilon )~~~
&\!\!\!\!=&
\int {{\cal D}  \eta }
\exp\left[ -\int _{t_a}^{t_b}dt \, \tilde H( \eta (t))\right]
\nonumber \\
&\!\!\!\!\times &\left[
-\epsilon c_1\partial _x
+\left(\epsilon c_2 + \epsilon ^2c_1\right)\frac{1}{2} \partial^2 _x
- \left(\epsilon c_3+ 3\epsilon ^2c_2c_1\right)\frac{1}{3!}\partial _x^3
\right.
\label{@18spiLPgl1w}
 \\
&\!\!\!\!+ &\left.
( \epsilon c_4+ \epsilon ^24c_3c_1+ \epsilon ^23c_2^2
+ \epsilon ^3c_2c_1^2+ \epsilon ^4c_1^2)
\frac{1}{4!}\partial ^4_x
+\dots\right]
 \psi
\left( x-\int_{t_a}^t dt' \eta(t')\right)
.   \nonumber
\end{eqnarray}
In the limit $ \epsilon \rightarrow 0$, only the linear terms
in $ \epsilon $ contribute,
which are all due
to the connected
parts of the correlation functions of $ \eta (t)$.
The differential operators in the brackets
can now be pulled out of the integral and we find
 the differential equation
\begin{eqnarray}
~\!\!\!\!\!\!\!\partial _t\psi(x,t )
&=&
\left[- c_1\partial _x
+ c_2 \frac{1}{2} \partial^2 _x
- c_3 \frac{1}{3!} \partial^3 _x
+c_4
\frac{1}{4!}\partial ^4_x+\dots\right]
 \psi
\left( x,t\right)
.
\label{@18spiLPgl1w3}
\end{eqnarray}
 We now
 replace
$c_1\rightarrow r_{x}$
 and
express
using
 (\ref{@cumuatsHa})
 the
differential operators in  brackets
as Hamiltonian operator
$-H_{r_{x}}(-i\partial _x)$. This leads to
 the Schr\"odinger-like equation
\begin{eqnarray}
~~~~~\partial _t\psi(x,t )
&=&
-H_{r_{x}}(-i\partial _x)
\,\psi\left( x,t\right)
.
\label{@18spiLPgl1w3H}
\end{eqnarray}

Due to the many derivatives in $H(i\partial _x)$, this equation is
in general  non-local.
This can be made explicit
with the help of the  L\'evy-Khintchine weight function $F(x)$
in the Fourier representation (\ref{@LevyKhich}).
In this case, the right-hand side
\begin{equation}
-H(i\partial _x)\psi(x,t)=
\int dx'\,
e^{-x'\partial _x}\,
F(x')\psi(x,t)
=\int dx'\,F(x')\psi(x-x',t)
.
\label{@LevyKhich1}\end{equation}
and the
Fokker-Planck-like equation
(\ref{@18spiLPgl1w3H})
takes the form of an integral equation.
Some people like to use  the
subtracted form (\ref{@LKSUB}) of the L\'evy-Khintchine
 and arrive at the
integro-differential equation
\begin{equation}
~~~~~\partial _t\psi(x,t )=
 \left[-c_1\partial _x -\frac{c_2}{2}\partial _x^2\right] \psi(x,t )+
\int dx'\,\bar F(x')\psi(x-x',t)
.
\label{@LevyKhich2}\end{equation}
The integral term can then be treated as
a perturbation to an ordinary Fokker-Planck equation.

By a similar procedure  as in the derivation of Eq.~(\ref{@18spiLPgl1w3})
it is possible
to derive a generalization
of Ito's rule (\ref{@Itosrule})
to
 functions of noise variable with non-Gaussian
distributions.
As in (\ref{@Itonew0})
we expand
$f( x(t+ \epsilon ))$:
\begin{eqnarray} ~ \!\!\!\!\!\!\!\!\!\!\!
&&~~\!\!\!\!\!\!\!\!\!
\!\!\!\!\!\!\!\!\!\!
f( x(t+ \epsilon ))\!=\!f(x(t)) +
\,f'(x(t))
 \int _t^{t+ \epsilon } dt'\,
 \dot x(t')
\nonumber \\&& ~~~~~~~~~~~~~ ~ ~~~~~ \hspace{1pt}
+~\frac{1}{2}f''(x(t))
\int _t^{t+ \epsilon } dt_1\int _t^{t+ \epsilon }
dt_2\,
 \dot x(t_1)
 \dot x(t_2)
\label{@Itonew20} \\&& ~~~~~~~~~~~~~~~~~~~\hspace{-1pt}
 \,
+~\frac{1}{3!}f^{(3)}(x(t))\int _t^{t+ \epsilon }
dt_1\int _t^{t+ \epsilon }
dt_2\int _t^{t+ \epsilon }
t_3\,\dot x(t_1)\dot x(t_2)\dot x(t_3)
+\dots~  ,  \nonumber
\end{eqnarray}
where $\dot x(t)= \eta (t)$
is the stochastic differential equation
with a nonzero expectation value $\langle  \eta (t)\rangle =c_1$.
In contrast to (\ref{@Itonew0})
which had to be carried out only up to second order
in $\dot x$,
we must now keep  {\em all\/} orders in the noise variable.
Evaluating the noise averages of the multiple integrals on the right-hand side
using the correlation functions
(\ref{@correlfu1})--(\ref{@correlfu}),
we find
 the
time derivative of the expectation value of an arbitrary function
of the fluctuating variable
$x(t)$
\begin{eqnarray}\!\!\!\!\!\!\!\!\!\!\!\!\!\!\!\! \!\!
 \!\langle f( x(t\!+\! \epsilon )) \rangle
\! &=&\!
\langle  f(x(t))\rangle +
\langle f'(x(t)) \rangle \epsilon c_1
+\frac{1}{2}\langle f''(x(t))\rangle ( \epsilon c_2\!+\! \epsilon ^2c_1^2)\nonumber \\
\! &\!\!\!\!\!\!\!\!\!\!\!\!\!\!\!\!\!&\!
~~~~\,~~~~~~~~\!+\frac{1}{3!}\langle f^{(3)}(x(t))\rangle
( \epsilon c_3
+ \epsilon ^2c_2c_1+
 \epsilon ^3c_1^3)
+\dots  \nonumber \\
&\!\!\!\hspace{1pt}=\!\!\!& \epsilon \left[- c_1\partial _x
+ c_2 \frac{1}{2} \partial^2 _x
- c_3 \frac{1}{3!} \partial^3 _x
+\dots\right]\langle f(x(t))\rangle+{\cal O}( \epsilon ^2).
\label{@Itonew0x}\end{eqnarray}
After the replacement
$c_1\rightarrow r_{x}$
the function
$f(x(t))$ obeys
therefore
 the following equation:
\begin{eqnarray}
 \langle \dot f( x(t))\rangle =  -H_{r_{x}}(i\partial _x)\langle f(x(t))\rangle .
\label{@Itonewex}\label{@Itonew}\end{eqnarray}
Taking out the lowest-derivative term this takes a form
\begin{eqnarray}
 \langle \dot f( x(t))\rangle = \langle  \partial _xf(x(t))\rangle
\langle \dot x(t)\rangle
-\bar H_{r_{x}}(i\partial _x)\langle f(x(t))\rangle .
\label{@Itonewbar}\end{eqnarray}
In  postpoint time slicing, this may be viewed as the expectation value of
the stochastic differential equation
\begin{eqnarray}
 \dot f( x(t)) =  \partial _xf(x(t))
\dot x(t)
-\bar H_{r_{x}}(i\partial _x)f(x(t)) .
\label{@Itonewex}\label{@Itonewb}\end{eqnarray}
This is the direct generalization of
 \ind{Ito's rule} (\ref{@Itosrule}).
\comment{and this may
be thought of a the expectation value of the
 stochastic equation
\begin{eqnarray}
 \dot f( x) = f'(x)\dot x -\bar H(i\partial _x)f(x),
\label{@Itonew}\end{eqnarray}
generalizing
Ito's
rule (\ref{@Itosrule}). }

For an exponential  function
$f(x)=e^{Px}$,
this becomes 
\begin{eqnarray}
 \frac{d}{dt}  e^{Px(t)} =\left[ \dot x(t) -H_{r_{x}}(i P)\right]e^{Px(t)}
 .
\label{@}\end{eqnarray}
\comment{a result closely related to  (\ref{@modwick}).}

\comment{Going back to the original
stochastic differential equation
 (\ref{@stdeqet})
in which
the average linear growth $r_{x}$
is shown explicitly
as
\begin{equation}
\dot x(t)=r_{x}+  \eta (t),
\label{@ratexvn}\end{equation}
with a vanishing noise average,
the relation (\ref{@Itonewex})
becomes
\begin{eqnarray}
\langle  \dot f( x)\rangle =
r_{x_S}\langle f'(x)\rangle
-\bar H(i\partial _x)\langle f(x)\rangle .
\label{@Itonewn}\end{eqnarray}
}
As a consequence of this equation for $P=1$,
the rate $r_S$ with which a
stock price
$S(t)=e^{x(t)}$
grows according to formula
(\ref{@stochdequsto}) is
now related to
$r_{x_S}$   by
\begin{equation}
r_S=
r_{x_S}-\bar H(i)=
r_{x_S}-[H(i)-iH'(0)]=
-H_{r_{x_S}}(i),
\label{@ratexvnru}\end{equation}
which replaces
  the simple
Ito relation
$r_{S}=r_{x_S}+\sigma ^2/2$ in Eq.~(\ref{@itoreln}).
Recall
the definition
$\bar H(p)\equiv H(p)-H'(0)p$
in Eq.~(\ref{@defhbar}).
The corresponding  generalization of  the left-hand part of
Eq.~(\ref{@stochdequstox})
 reads
\begin{equation}
\frac{\dot S}{S}
=\dot x(t)
-\bar H(i)
=
\dot x(t)
-[H(i)-iH'(0)]
=\dot x(t) -r_{x_S}
-H_{r_{x_S}}(i)
.
\label{@stochdequstoxn}\end{equation}
The forward price of a stock must therefore be calculated with
the generalization of formula
(\ref{@forwS}):
\begin{equation}\!\!
\langle S(t)\rangle =S(0)e^{r_St}=
S(0)\langle e^{r_{x_S}t+\int_0^t dt'\, \eta (t')}\rangle
=S(0)e^{-H_{r_{x_S}}(i)t}
=S(0)e^{\{r_{x_S}-[H(i)-iH'(0)]t\}}.
\label{@forwSn}\end{equation}

Note that
may derive
the differential equation of an arbitrary function $f(x(t))$
in Eq.~(\ref{@Itonew}) from a simple
mnemonic rule, expanding sloppily
\begin{eqnarray}  \!\!\!\!\!\!\!\!\!\!\!\!\!\!\!\!\!\!\!\!
f(x(t+ dt ))&=&
f(x(t)+\dot x dt)=
f(x(t))+f'(x(t))\dot x(t) dt+
\frac{1}{2}f''(x(t))\dot x^2(t)dt^2
\nonumber \\&\!\!\!+\!\!\!&
\frac{1}{3!}f^{(3)}(x(t))\dot x^3(t)dt^3
+\dots~,
\label{@20.364n}\end{eqnarray}
and replacing
(\ref{@Itosrule}),
\begin{eqnarray}  \!\!\!\!\!\!\!\!\!\!
\langle\dot x(t)\rangle dt
\rightarrow c_1 dt,~~ ~
\langle\dot x^2(t)\rangle dt^2
\rightarrow c_2 dt,~~  ~
\langle\dot x^3(t)\rangle dt^2
\rightarrow c_3 dt,\dots\,.~~~~
\label{@Itosruleg}\end{eqnarray}

\section{Martingales}
\label{Martingales}
In financial mathematics,
an often-encountered  concept is that of a \iem{martingale} \cite{Mar}.
The name stems from a casino
 strategy
in which a gambler doubles his stake
each time a bet
is lost.
A stochastic variable
is called a martingale,
if its expectation value
 is time-independent.
The noise variable $ \eta (t)$
with vanishing average is a trivial martingale.

\subsection{Gaussian Martingale}
For a harmonic noise variable,
the exponential $e^{\int_0^t dt' \eta (t')- \sigma ^2t/2}$
is a nontrivial
martingale, due to Eq.~(\ref{@forwS}). For the same reason, a
 stock price
$S(t)=e^{x(t)}$
with
$x(t)$ obeying the stochastic differential equation
 (\ref{@stochdequstox})
can be made a martingale
by a time-dependent multiplicative factor
the quantity
$e^{-r_St}e ^{\dot x }=e^{-r_{x_S}t}e ^{\dot x - \sigma ^2t/2}$.
An explicit
distribution which makes
$S(t)=e^{x(t)}$
a martingale is
\begin{eqnarray}
P^M (x_b t_b|x_at_a)\equiv \frac{e^{-r_St}}{ \sqrt{ 2\pi \sigma^2(t_b-t_a)}}
\exp\left\{-
 \frac{[x_b-x_a- r_{x_S}(t_b-t_a)]^2}
{2 \sigma^2(t_b-t_a)}
\right\}                     .
\label{@martinga0}\end{eqnarray}
I can easily be verified by direct integration
that
the expectation
value  of
$S(t)=e^{x(t)}$
is time-independent
with this distribution:
\begin{eqnarray}
\!\!\!\langle S(t_b)\rangle =
\langle e^{x(t_b)}\rangle =
\int dx_b\,
e^{x_b}
 \,
 P^M (x_b t_b|x_at_a)
\label{@20meatrco}\end{eqnarray}
is independent of the time $t_b$.

At this place
we can make an important observation:
There exists an entire family of distributions
for which $S(t)=e^{x(t)}$
is a martingale, namely
\begin{eqnarray}
P^{M\hspace{1pt}r} (x_b t_b|x_at_a)\equiv \frac{e^{-rt}}{ \sqrt{ 2\pi \sigma^2(t_b-t_a)}}
\exp\left\{-
 \frac{[x_b-x_a- r_{x}(t_b-t_a)]^2}
{2 \sigma^2(t_b-t_a)}
\right\}                     .
\label{@martinga1}\end{eqnarray}
for any $r$ and $r_x=r- \sigma ^2/2$.
Such distributions which differ  by the
drift $r$ are called  {\em equivalent\/}.
The prefactor $e^{-r t}$
is  referred to as a {\em discount factor\/} with the rate
$r $.
\iems{discount+factor}

\subsection{Non-Gaussian Martingales}

For $S(t)=e^{x(t)}$ with an arbitrary non-Gaussian noise
$ \eta (t)$, there are many ways of constructing  martingales.
the relation (\ref{@ratexvnru})
allows us to construct  immediately the simplest martingale

\subsubsection{Natural Martingale}
An obvious generalization
of the Gaussian expression   $e^{\int_0^t dt' \eta (t')- \sigma ^2t/2}$
is the exponential
$
e^{\int_0^t dt' \eta (t')+\bar H(i)t}$,
 whose expectation value is time-independent due to
Eq.~(\ref{@ratexvnru}).
Expressed in terms of
 the associated stock price
$S(t)=e^{x(t)}$, we obtain the simplest martingale
\begin{equation}
e^{-r_St}
S(t)=
e^{-r_St}
e^{r_{x_S}t+\int_0^t dt' \eta (t')}
\label{@}\end{equation}
if $r_S$ and $r_{x_S}$  are related by
$r_S=
r_{x_S}-\bar H(i)
=- H_{x_S}(i)$.

It is easy to write down a
distribution function which makes $S(t)=e^{x(t)}$ itself a martingale:
\begin{eqnarray}
\!\!\!\!\!\!\!\!\!
P^M(x_b t_b|x_at_a)=e^{-r_St}
\int {{\cal D}  \eta }
\int {{\cal D}  x}
\exp\left\{- \int _{t_a}^{t_b}dt \,
 \tilde H_{r_{x_S}}( \eta (t))\right\}
 \delta [\dot x-  \eta ].
\label{@18spiLPEM}
	  \end{eqnarray}
There exists also here
an entire  family of equivalent distribution functions
for which
the integrals of the type (\ref{@20meatrco}) over $e^{x_b}$
are independent of $t_b$. This is done as follows.
One obvious family of this type is a straightforward
generalization of the Gaussian family (\ref{@martinga1})
which we shall refer to as
{\iems{natural martingale}}
{\em{natural martingales\/}}:
\begin{eqnarray}
\!\!\!\!\!\!\!\!\!
P^{M\hspace{1pt}r}(x_b t_b|x_at_a)=e^{-rt}
\int {{\cal D}  \eta }
\int {{\cal D}  x}
\exp\left\{- \int _{t_a}^{t_b}dt \,
 \tilde H_{r_{x}}( \eta (t))\right\}
 \delta [\dot x-  \eta ],
\label{@18spiLPEMge}
	  \end{eqnarray}
with arbitrary $r=r_x-\bar H(i)=-H_{r_x}(i)$.
Indeed, multiplying this with $e^{x_b}$ and integrating over $x_b$
gives rise to  a $ \delta $-function  $ \delta (p-i)$ and produces  the
same result
$e^{x_a}$ for all times $t_b$.
The path integral is solved
 by the  Fourier integral    [compare (\ref{@18spiLP3})]
\begin{eqnarray}
\!
\!
\!
\!
P^{M\hspace{1pt}r}(x_b t_b|x_at_a)=
e^{-rt}
\int_{-\infty } ^\infty  \frac{{d}  p }{2\pi}
\exp\left[ ip(x_b-x_a)-({t_b}-{t_a}) H_{r_x}(  p) \right]
\label{@18spiLP3p1}      .
\end{eqnarray}

\subsubsection{Esscher Martingale}

In the literature on mathematical finance, much attention is given
another  family of
equivalent martingale measures.
It has been used a long time ago to estimate risks of actuaries \cite{Ess}
and introduced more recently into the theory
of option prices \cite{GerberS,HP}
where it is now of wide use
\cite{ch2013}--\cite{raible}.
This family is constructed as follows.
Let $\tilde D(x)$ be an arbitrary distribution function
with a Fourier transform
\begin{equation}
\tilde D(x)=
 \int _\infty^\infty\frac{dp}{2\pi}\,e^{-H(p)}\,e^{ipx},
\label{@}\end{equation}
and
 $H(0)=0$, to guarantee a
 unit normalization $\int dx\tilde D(x)=1$.
We now introduce an
{\em{Esscher-transformed}}\iems{Esscher+transform}
distribution function. It is obtained by slightly tilting
the initial distribution $\tilde D(x)$
by multiplication with an asymmetric
exponential factor $e^{\theta x}$:
\begin{equation}
D^\theta(x)=e^{H(i\theta)}~ e^{\theta x}\tilde D(x).
\label{@tiltedD}\end{equation}
The constant prefactor $e^{H(i\theta)}$
is necessary to conserve the total  probability.
This distribution
can be written as a
Fourier transform
\begin{equation}
D^\theta(x)=
 \int _\infty^\infty\frac{dp}{2\pi}\,e^{-H^\theta(p)}\,e^{ipx}.
\label{@18EssTr}\end{equation}
with the Esscher-transformed Hamiltonian
\begin{equation}
 H^\theta(p)
 \equiv H(p+i\theta)
 - H(i\theta).
\label{@EssHam}\end{equation}
Since $H^\theta(0)=0$,
the transformed distribution is properly normalized
$\int dxD^\theta(x)=1$.
 We now define the Esscher-transformed  expectation value
\begin{equation}
\langle F( x)\rangle ^\theta\equiv
\int dx D^\theta(x)F(x).
\label{@}\end{equation}
It is related to the original expectation value by
\begin{equation}
\langle F( x)\rangle ^\theta\equiv e^{H(i\theta)}\,\langle e^{\theta x} F(x)\rangle
\label{@20Theta}\end{equation}
For the specific function $F(x)=e^x$,
Eq.~(\ref{@20Theta}) becomes
\begin{equation}
\langle e^x\rangle ^\theta=
 e^{-H^\theta(i)}
\equiv
e^{H^\theta(i\theta)}\,\langle e^{(\theta+1) x} \rangle
 =
e^{H^\theta(i\theta)
-H^\theta(i\theta+i)}.
\label{@20Theta2}\end{equation}

Applying the transformation
(\ref{@18EssTr}) to each time slice in the general path integral
(\ref{@18spiLP}),
we obtain the
{\em{Esscher-transformed}}\iems{Esscher+transform}
path integral
\begin{eqnarray}
&&\!\!\!\!\!\!\!\!\!\!\!P^{\theta}(x_b t_b|x_at_a)\!=\!e^{-r_St}
e^{H_{r_{x_S}}(i\theta)t}
\int \!{{\cal D}  \eta }
\int \!{{\cal D}  x}
\exp\left\{ \int _{t_a}^{t_b}dt \,
\left[ \theta  \eta (t)- \tilde H_{r_{x_S}}( \eta (t))\right] \right\}
 \delta [\dot x-  \eta ],   \nonumber \\&&
\label{@18spiLPE}
\end{eqnarray}
The solution is given by the  Fourier integral    [compare (\ref{@18spiLP3p1})]
\begin{eqnarray}
&&\! \!\!\!\!\!\!\!\!\!\!\!\!\!
\!
\!
\!
P^\theta(x_b t_b|x_at_a)= e^{-r_St}
e^{H_{r_{x_S}}(i\theta)t}
\int_{-\infty } ^\infty  \frac{{d}  p }{2\pi}
\exp\left[ ip(x_b-x_a)-({t_b}-{t_a}) H_{r_{x_S}}(  p+i\theta) \right]
\label{@18spiLP3p}      .\nonumber \\&&
\end{eqnarray}

Let us denote the expectation values calculated with this probability
by $
\langle ~\dots~\rangle ^\theta$.
Then we
 find
for $S(t)=e^{x(t)}$  the time dependence
\begin{equation}
\langle S(t)\rangle ^\theta=
e^{-H^\theta_{r_{x_S}}(i)t}.
\label{@20Esstrav}\end{equation}
This equation shows, that
the exponential of a
stochastic variable $x(t)$
can be made a martingale with respect to any
Esscher-transformed distribution
if we remove the
 exponentially growing factor
$\exp({r^\theta t})$
with
\begin{equation}
r^\theta\equiv  - H_{r_{x_S}}^\theta(i)=
-H_{r_{x_S}}(i+i\theta)
+H_{r_{x_S}}(i\theta).
\label{@choiceEs}\end{equation}
Thus, a family of  equivalent martingale distributions
for the stock price
 $S(t)=e^{x(t)}$
is
\begin{equation}
P^{ M\hspace{1pt}\theta}(x_b t_b|x_at_a)\equiv e^{-r_\theta t}P^{\theta}(x_b t_b|x_at_a)
\label{@}\end{equation}
for any
choice of the  parameter $\theta$.
\comment
{We may also define an
{\em{Esscher-transformed}}\iems{Esscher+transform}
fluctuating stock price, where the constant prefactor in (\ref{@tiltedD})
grows now linearly in time,
\begin{equation}
S^\theta(t)\equiv e^{H_{r_{x_S}}(i\theta)t} ~
 e^{\theta x(t)}
 S(t).
\label{@Esschertrst}\end{equation}
Its ordinary expectation value is equal to
(\ref{@20Esstrav}):
\begin{equation}
\langle S^\theta(t)\rangle =
\langle S(t)\rangle ^\theta =e^{-H_{r_{x_S}}^\theta(i)t}.
\label{@}\end{equation}
It is now obvious that
with the choice (\ref{@choiceEs}) of the Esscher parameter $\theta$,
the transformed stochastic variable
\begin{equation}
S^D^\theta(t) \equiv
 e^{-rt}S^\theta(t)=
 e^{-rt}e^{H_{r_{x_S}}(i\theta)t}
e^{\theta x(t)}S(t)
= e^{H_{r_{x_S}}(i\theta+i)t}
e^{(\theta+1) x(t)}
\label{@EsstrS}\end{equation}
has a time-independent expectation value
due to Eq.~(\ref{@20Theta2}).
The Esscher-transformed distribution
with the Hamiltonian $H_{r_{x_S}}^\theta(p)$
is called \iem{risk-neutral} distribution.
}

For a harmonic distribution function (\ref{@martinga0}),
the Esscher martingales and the previous ones are equivalent.
Indeed, starting from (\ref{@martinga0}) in which
$r_S=r_{x_S}+ \sigma ^2/2$,
the Esscher transform leads us after a quadratic completion to
the family of natural martingales
(\ref{@martinga0})  with the rate parameter
$r=r_{x_S}+\theta  \sigma ^2$.

\subsubsection{Other Non-Gaussian Martingales}
Many other non-gaussian martingales
have been discussed in the literature.
Mathematicians have invented various sophisticated
criteria under which
one would be preferable over the others
for calculating financial risks.
Davis has introduced a so-called \iem{utility function}
\cite{Davis} which is supposed to select optimal martingales
for different purposes.
\comment{There exists also a so-called minimal martingale
\cite{FS}.  but the mathematical setup
in these discussions is hard to understand. }

For the upcoming development of a theory of option pricing,
 only the initial {\em natural martingale\/}
will turn out to be relevant.

\section{Option Pricing}

The  most important
use of path integrals
in financial markets is made
in the determination
of
a fair  price of financial derivatives, in particular options.
Options are an ancient
financial tool. They are
used for speculative purposes or for hedging\ins{hedging}
major market transactions against unexpected changes
in the market environment. These
can sometimes
produce dramatic price explosions
or erosions, and options are supposed to
prevent the destruction of huge amounts of capital.
 Ancient Romans, Grecians, and Phoenicians traded options
against outgoing cargos from their local seaports.
In financial markets,
 options are contracted between two parties
in which one party has the right but not the
obligation to do something, usually to buy or sell some underlying asset.
Having rights without obligations has a
financial value,
so option holders must
pay a price for acquiring them.
The price depends on the value
 of the associated asset, which is why they are also called
 {\em derivative assets\/}\iems{derivative assets}
or briefly {\em derivatives\/}.
{\em Call options\/}\iems{call options}\iems{options,call} are
contracts giving the option holder the right to buy something, while
 {\em put options\/}\iems{put options}\iems{options,put}
entitle the holder to sell something.
The price of an
option is called {\em premium\/}.\iems{premium}
Usually, options are associated with stock,
bonds, or commodities like oil, metals or other raw materials.
In the sequel we shall consider call options on stocks,
to be specific.

Modern option pricing techniques
have their roots in
early work
by Charles Castelli\aus{C. Castelli}
who published
 in 1877
 a book entitled
{\em The Theory of Options in Stocks and Shares\/}.
 This book presented an introduction to the hedging
and speculation aspects of options. However, it still
lacked a sound theoretical base. Twenty three years later,
Louis Bachelier\aus{L. Bachelier} offered the earliest known analytical valuation for options
in his dissertation at
the Sorbonne \cite{LB}.
It is curious that Bachelier
discovered the treatment of stochastic
 phenomena five years before Einstein's related
 but much more
famous work on Brownian motion \cite{Einstein}, and 23 years before Wiener's
mathematical development \cite{Wiener}.
The stochastic differential equations
considered by him
still
had an important defect
of allowing for
negative security prices, and for option prices
 exceeding the price of the underlying asset.
Bachelier's work was continued
by
Paul Samuelson\aus{P. Samuelson}, who wrote in 1955
 an unpublished paper entitled
{\em Brownian Motion in the Stock Market\/}.
During that same year,
Richard Kruizenga,\aus{R. Kruizenga,}
 one
of Samuelson's students,
cited Bachelier's work in his dissertation {\em Put and Call Options:
A Theoretical and Market Analysis\/}. In 1962, another
dissertation, this time by A. James Boness\aus{A.J. Boness},
focused on options. In his work,
entitled {\em A Theory and Measurement of Stock Option Value\/},
Boness
developed a pricing model that made a significant theoretical jump
from that of his predecessors.
More significantly, his work served as a precursor to
that of Fischer Black\aus{F.Black} and Myron Scholes\aus{M. Scholes},
who in 1973 introduced their famous
\iem{Black and Scholes Model} \cite{BS}
which, together with the
improvements introduced by R. Merton\aus{R. Merton},
earned the Nobel prize in 1997.\footnote{For F. Black
the prize came too late---he had died two years earlier.}

\subsection{Black-Scholes Option Pricing  Model}

In the early seventies,
 Fisher Black was working
on a valuation model for stock
warrants and observed that
his formulas
 resembled
very much the
 well-known equations for heat transfer.
Soon after this,
Myron Scholes joined Black and together they
discovered
 an approximate
 option pricing model which is still of wide use.
It is
an improved
version of a previous
model developed by
 A. James Boness in his Ph.D. thesis the University of Chicago.

The Black and Scholes Model is based on the following assumptions:
\begin{enumerate}
\item  The logarithm of the returns is normally distributed.
We remarked before
that
there are considerable deviations
which  call for
improvement
of the model to be developed below.

\item  Markets are efficient.
\ins{efficient+markets}\ins{markets,efficient}%
This assumption implies that
the market operates continuously with
share prices following a continuous stochastic
process without memory.
It also implies that
different markets have the same asset prices.

This is not quite true.
Different markets
do in general have slightly different prices.
Their differences are kept small
by the existence
of
arbitrage dealers\ins{arbitrage dealers}.
There also
 exist correlations
over a short time scale
which make it possible, in principle,
to profit without risk from
statistical arbitrage.
This possibility
is, however,
 strongly
limited
by transaction fees.

\item  No commissions are charged.\ins{commissions}

This is not true.
Usually market participants have to pay a commission to buy or sell assets.
Even floor traders pay some kind of fee, although this
is usually very small. The
fees payed by individual investors
 is more substantial and can distort the output of the model.

\item Interest rates remain constant and known.

The Black and Scholes model assumes the existence of a
risk-free rate to represent this constant and known rate.
In reality there is no such thing as the risk-free rate.
As an approximation, one uses
the discount rate on U.S. Government Treasury Bills with 30 days left
until maturity. During periods of rapidly changing
interest rates, these 30 day rates are often subject to change,
thereby violating one of the assumptions of the model.

\item The stock pays no dividends during the option's life.\ind{dividends}

Most companies pay dividends to their share holders, so this
is
a limitation to the model
since
 higher dividend
lead to lower call premiums.
There is, however, a simple
possibility of adjusting the model
to the real  situation
by  subtracting the discounted value of a future dividend from the
stock price.

\item European exercise terms are used.
\ins{European+option}\ins{option,European}
European exercise terms
imply the
exercise of an option
only
on the expiration date. This is in contrast
to the American exercise terms
which
allow for this
at any time during the life of the option. This
greater flexibility
makes
an  American option more valuable
than the European one.
\ins{American+option}\ins{option,American}

The difference is, however, not dramatic
 in praxis
 because\ins{call}
very few calls are ever exercised before the last few days of their life,
since
an early  exercise
means giving away the remaining time
value on the call.
Different exercise times towards the end of the life of a call
are irrelevant since
the remaining time value is very small
and  the intrinsic value has a small  time dependence,
barring a dramatic event right before expiration date.

\end{enumerate}

Since 1973, the original Black and Scholes Option Pricing Model has been the subject of much attention.
In the same year, Robert Merton\aus{R. Merton} \cite{Mer}
 included the effect of dividends. Three years later,
Jonathan Ingerson\aus{J. Ingerson}
 relaxed the
assumption of no taxes or transaction costs,
and Merton  removed the restriction of constant interest rates.
At present we are in
a position of being able to determine
quite reliably the values of stock options.

The
relevance of path integrals
to this field was recognized first
in
1988 by a theoretical physicist \aut{J.W. Dash},
who wrote two unpublished papers on the subject
entitled
{\em Path Integrals and Options I} and {\em II\/} \cite{Dash}.
Since then many theoretical physicists
have entered the field, and papers on this subject
have begun appearing on the Los Alamos server \cite{BP,FM,Otto}.

\subsection{Evolution Equations of Portfolios with Options}

The option price $O(t)$ has a larger fluctuations
than the associated stock price.  It usually varies with an
efficiency
factor
$\partial O(S(t),t)/\partial S(t)$.
For this reason it is possible, in the ideal case of Gaussian
price fluctuations, to guarantee a steady growth
of a portfolio
by mixing $N_S(t)$ stocks with $N_O(t)$  options and
a certain amount of cash which is usually kept in the form of bonds,
whose number is denoted by $N_B(t)$.
The composition $[N_S(t),N_O(t),N_B(t)]$ is referred to as the
\iem{strategy} of the portfolio manager.
The total wealth has the
 value
\begin{equation}
W(t)=N_S(t)S(t)+N_O(t)O(S,t)+N_B(t)B(t).
\label{@totalprot}\end{equation}
The goal is to make it grow
with a smooth exponential curve {\em without fluctuations\/}
\begin{equation}
\dot W(t)\approx r_WW(t).
\label{@18vdot}\end{equation}
As we shall see immediately, this is possible provided
the short-term  bonds (usually those
with 30 days to maturity)
grow without fluctuations.
\begin{equation}
\dot B(t)\approx r_BB(t).
\label{@18vdotB}\end{equation}
Under this assumption,
the rate $r_B$ is referred to as
{\em riskfree interest rate\/}%
\iems{riskfree+interest+rate}\iems{rate,riskfree}
This assumption
 is fulfilled in
true markets only approximately
since there can always be events
which change the value of short-term bonds excessively.

The existence of arbitrage dealers
will ensure that the growth rate $r_W$
is equal to
that
of the short-term  bonds
\begin{equation}
r_W\approx r_B.
\label{@rVB}\end{equation}
Otherwise the dealers would
change from one investment to the
other.

In the decomposition
(\ref{@totalprot}), the desired growth (\ref{@18vdot})
reads
\begin{eqnarray}
 &&
\!\!\!\!\!\!\!\!\!
\!\!\!\!\!\!\!\!\!\!\!
N_S(t)\dot S(t)+N_O(t)\dot O(S,t)+  N_B(t)\dot B(t)  +
\dot  N_S(t)S(t)+\dot N_O(t)O(S,t)+\dot N_B(t) B(t)\nonumber \\
&&~~~~~~=r_W\left[N_S(t)S(t)+N_O(t)O(S,t)+N_B(t)B(t) \right] .
\label{@}\end{eqnarray}
Due to (\ref{@18vdotB}) and
(\ref{@rVB}), the terms containing $N_B(t)$
without a dot  drop out.
Moreover, if no extra money is inserted into or
taken from the system, i.e., if
 stocks, options, and bonds are only
traded against each other,
this does  not change the total wealth,
assuming the absence of commissions.
This
 so-called {\iem{self-financing strategy}}
is expressed in the
equation
\begin{eqnarray}
 \dot  N_S(t)S(t)+\dot N_O(t)O(S,t)+\dot N_B(t) B(t) =0.
\label{@}\end{eqnarray}
Thus the growth equation (\ref{@18vdot})
translates into
\begin{eqnarray}
\dot W(t)=N_S\dot S+ N_O\dot O+ N_B\dot B= r_W\left( N_S S+N_O O
+N_B\dot B \right)      .
\label{@grequ0}\end{eqnarray}
Due to the equality
of
the rates
$r_W=r_B$
and Eq.~(\ref{@18vdotB}),
  the entire contribution of
 $B(t)$ cancels, and we obtain
\begin{eqnarray}
N_S\dot S+ N_O\dot O= r_W\left( N_S S+N_O O
\right)
\label{@Glrlinks0}\end{eqnarray}
The important observation is now that there
exists an optimal ratio
between the number of stocks
 $N_S$ and the number of options
 $N_O$,
which is inversely equal to the
efficiency
 factor
\begin{equation}
\frac{N_S(t)}{N_O(t)}=-\frac{\partial O(S(t),t)}{\partial S(t)}.
\label{@18dcds}\end{equation}
 Then Eq.~(\ref{@Glrlinks0}) becomes
\begin{eqnarray}
N_S\dot S+ N_O\dot O=
r_W\left(-\frac{\partial O}{\partial x}+O\right)N_O
 .
\label{@Glrlinks}\end{eqnarray}
The two terms on the left-hand side are treated  as follows:
First we use the relation
(\ref{@18dcds})
to
rewrite
\begin{eqnarray}
N_S\dot S  =-\frac{\partial O(S,t)}{\partial S}\dot S
 =-\frac{\partial O(S,t)}{\partial x}\frac{\dot S}{S}
,
\label{@}\end{eqnarray}
and further, with the help of
Eq.~(\ref{@stochdequstoxn}), as
\begin{eqnarray}
N_S\dot S  =
 -\frac{\partial O(S,t)}{\partial x}\left[ \dot x-\bar H(i)\right]
.
\label{@20NSdot}\end{eqnarray}
In the second term on the left-hand side of (\ref{@Glrlinks}),
we expand
the total time dependence of
the option price in a Taylor series
\begin{eqnarray}
\frac{d O}{dt}&=&\frac{1}{dt}
\Big[ O(x(t)+\dot x(t)\,dt,t+dt)
-O(x(t),t)\Big]
\nonumber \\&=&
\frac{\partial O}{\partial t}  +
\frac{\partial O}{\partial x }\, \dot x+
\frac{1}2 \frac{\partial^2 O}{\partial x^2}\,\dot x^2\,dt
+\frac{1}{3!} \frac{\partial^3 O}{\partial x^3}\,\dot x^3\,
dt^2
+\dots~.
\label{@20expanC}\end{eqnarray}
We we have
gone over to the logarithmic
stock price
variable $x(t)$ rather than
$S(t)$ itself.
Some of the derivatives on the right-hand side are denoted
by special symbols
in financial mathematics:
the quantities
$\vartheta\equiv \partial O/\partial t$,
$ \Delta \equiv
\partial O/\partial S=
\partial O/\partial x S$,
 and
$ \Gamma \equiv \partial^2 O/\partial S^2=(\partial^2 O/\partial x^2-
\partial O/\partial x)/S^2$
are called the ``Theta",  ``Delta", and  ``Gamma" of the option.
Another derivative with a standard name is
the ``Vega" $V\equiv \partial O/\partial  \sigma $.

The expansion (\ref{@20expanC}) is
  carried to arbitrary powers of $\dot x$
as in
 (\ref{@20.364n}). It is, of course,
 only an abbreviated notation for the proper expansion
in powers of a stochastic variable
to be  performed as in
Eq.~(\ref{@Itonew20}).
Inserting (\ref{@20expanC})
and (\ref{@20NSdot}) on the left-hand side of
Eq.~(\ref{@Glrlinks}), this becomes
%
\begin{eqnarray}
N_S\dot S
+N_O\dot O
&=&-~N_O\frac{\partial O}{\partial x}
\left[\dot x-\bar H(i)\right]
\nonumber \\
&&+~
N_O\left(
 \frac{\partial O}{\partial t}  + \frac{\partial
O}{\partial x }\, \dot x+ \frac{1}2 \frac{\partial^2 O}{\partial
x^2}\dot x^2dt
+ \frac{1}{3!} \frac{\partial O}{\partial x}\dot x^3dt
+\dots
\right)\nonumber \\
&=&N_O
\left[ \bar H(i)
\frac{\partial O}{\partial x}+\frac{\partial O}{\partial t}+
 \frac{1}{2} \frac{\partial O}{\partial x}\dot x^2dt
+ \frac{1}{3!} \frac{\partial O}{\partial x}\dot x^3dt
+\dots\right].      ~~~~~~~~~~
\label{@substfurt}\end{eqnarray}
Remarkably, the fluctuating variable $\dot x$
drops out in this equation, which therefore looses its stochastic character.
At the same time, it becomes
 {\em independent\/}
of the growth rate $r_S$ of the stock price.
This is the reason
why the
total wealth
$ W(t)$
increases without fluctuations.

We now
treat the Taylor series
\begin{equation}
 \frac{1}{2} \frac{\partial ^2O}{\partial x^2}\dot x^2\,dt
+ \frac{1}{6} \frac{\partial ^3O}{\partial x^3}\dot x^3\,dt^2
\!+\!\dots
\label{@20Tay}\end{equation}
in the same way as the expansion
 (\ref{@20.364n}),
  using the rules (\ref{@Itosruleg}),
such that (\ref{@20Tay}) becomes
\begin{equation}
 \frac{1}{2} \frac{\partial ^2O}{\partial x^2}\dot x^2\,dt
+ \frac{1}{6} \frac{\partial ^3O}{\partial x^3}\dot x^3\,dt^2
\!+\!\dots
=-\bar H(i\partial _x)O .
\label{@}\end{equation}
In this way we find
for the option price
$O(x,t) $
the Fokker-Planck-like differential  equation
\begin{equation}
\frac{\partial  O }{\partial t}=r_WO
-r_{x_W}\frac{\partial O}{\partial x}
+\bar H(i\partial_x )O ,
\label{@FPlike}\end{equation}
where we have defined,
by analogy with
(\ref{@ratexvnru}), an auxiliary rate parameter
\begin{equation}
 r_{x_W}\equiv r_W    +\bar H(i)
.
\label{@ratexv}\end{equation}
Note, however, that in contrast
to the relation between $r_{x_S}$ and $r_S$
defined for a fluctuating stock price $S(t)$ and its logarithm $x(t)$,
the parameter $r_{x_W}$ {\em does not\/} have the physical interpretation
of governing the logarithm of $W(t)$
since   the absence of fluctuations in the wealth
 $W(t)$
makes  $\log W(t)$
 grow linearly
with the {\em same\/} rate $r_W$  that governs
the exponential growth
$e^{r_Wt}$ of
$W(t)$ itself.

If we rename $t$ as $t_a$,
The general solution
 of the
differential equation  (\ref{@FPlike}), which at some time
$t=t_b$
starts out like $ \delta (x-x_b)$,
has the Fourier representation
\begin{equation}
\!\!\!\!\!P(x_b t_b|x_at_a)=e^{-r_W(t_b-t_a)}
\int_{-\infty}^\infty
 \frac{dp}{2\pi}e^{ip(x_b-x_a)}\exp\left\{ -\left[\bar H(p)
+ir_{x_W}p\right] (t_b-t_a)\right\} ,
\label{@18stprgf}\end{equation}
if the {\em initial\/} variables $x_a$ and $t_a$ are identified with
$x$ and $t$, respectively.
A convergent integral exists only for $t_b>t_a$.

\comment{The discounted option price
$O^D(x,t)\equiv e^{-r_Wt}  O(x,t)$
satisfies the partial differential equation
\begin{equation}
\frac{\partial  O^D }{\partial t}=
-r_{x_W}\frac{\partial O^D}{\partial x}
+\bar H(i\partial_x )\,O^D =
 H_{r_{x_W}}(i\partial_x )\,O^D .
\label{@FPlikeD}\end{equation}
By analogy with (\ref{@Esschertrst}),
we may also define an  Esscher-transformed option price,
with a negative time in the prefactor
since the time evolution equation for $O(x,t)$
runs backwards in time:
\begin{equation}
O^D^\theta(x,t)\equiv e^{H_{r_{x_W}}(i\theta)t}
e^{\theta x} O^D^\theta(x,t).
\label{@}\end{equation}
Since
\begin{equation}
e^{-\theta x} H_{r_W}(i\partial_x ) e^{\theta x}
=H_{r_W}(i\partial_x+i\theta ),
\end{equation}
this satisfies the partial differential equation
\begin{equation}
\frac{\partial  O^D^\theta }{\partial t}=
 \left[ H_{r_{x_W}}(i\partial_x+i\theta )-{H_{r_{x_W}}(i\theta)}\right]
\,O^D .
\label{@FPlikeDth}\end{equation}
The function
$
O^D^\theta(x,t)$ is therefore driven by the Esscher-tranformed Hamiltonian
(\ref{@EssHam}) with $p$ replaced by $i\partial _x$.
}

Comparing this expression with
Eq.~(\ref{@18spiLP3p1}) we recognize
it as a member of the family of equivalent
martingale measures $P^{M\hspace{1pt}r}(x_b t_b|x_at_a)$,
in which the discount factor $r$ coincides with
the \ind{risk-free interest rate} $r_W$.

\subsection{Option Pricing for Gaussian  Fluctuations}

For Gaussian fluctuations where $H(p)= \sigma ^2p^2/2$,
the
 integral in (\ref{@18stprgf}) can easily be
performed
and
 yields
\begin{equation}
P(x_b t_b|x_at_a)=
\Theta(t_b-t_a)
\frac{e^{-r_W(t_b-t_a)}}{ \sqrt{ 2\pi \sigma^2(t_b-t_a)}}
\exp\left\{-
 \frac{[x_b-x_a-
r_{x_W}(t_b-t_a)]^2}
{2 \sigma^2(t_b-t_a)}
\right\} .
\label{@stprGF}\end{equation}
%
This probability distribution is  obviously
 the solution of the
 path integral
\begin{equation}
   P (x_b t_b|x_at_a)=
\Theta(t_b-t_a)
e^{-r_W(t_b-t_a)}
\int {\cal D}x\,\exp\left\{ -\frac{1}{2 \sigma^2}\int _{t_a}^{t_b}\left[ \dot x-r_{x_W}\right] ^2\right\} .
\label{@}\end{equation}

Recalling the discussion in Section~\ref{Martingales},
the distribution function (\ref{@stprGF})
is recognized as a member of the  equivalent
family of \ind{martingale} distributions (\ref{@martinga1}) for
the stock price $S(t)=e^{x(t)}$.
It is the particular distribution in which  the discount factor
contains the risk-free interest rate $r_W$, i.e.,
 (\ref{@stprGF}) is equal to
 the martingale distribution
$P^{M\hspace{1pt}r_W} (x_b t_b|x_at_a)$.
This  distribution
is referred to as the
 \iem{risk-neutral} equivalent martingale distribution.

An option is written for a certain \iem{strike price}
$E$ of the stock.
The value of the option at its expiration date $t^b$
is given by the difference between
the stock price on expiration date and the strike price:
\begin{equation}
O(x_b,t_b)=\Theta(S_b-E) (S_b-E)=\Theta(x_b-x_E)( e^{x_b}-e^{x_E}),
\label{@strikepr}\end{equation}
where
\begin{equation}
x_E\equiv \log E .
\label{@}\end{equation}
The Heaviside function accounts for the fact that
only for $S_b>E$ it is worthwhile to execute the option.

From (\ref{@strikepr}) we calculate
the option price at an arbitrary earlier time
using the time evolution amplitude (\ref{@stprGF})
\begin{equation}
O(x_a,t_a)=\int_{-\infty}^\infty dx_b\, O(x_b,t_b)\,P^{M\hspace{1pt}r_W} (x_b t_b|x_at_a).
\label{@BSINTEGRAL}\end{equation}
Inserting  (\ref{@strikepr})
we obtain the sum of  two terms
\begin{equation}
O(x_a,t_a)=
O_S(x_a,t_a)-
O_E(x_a,t_a),
\label{@optionpr}\end{equation}
where
\begin{eqnarray}
\!\!\!\!\!\!\!\!O_S(x_a,t_a)=
\frac{e^{-r_W(t_b-t_a)}}{ \sqrt{ 2\pi \sigma^2(t_b-t_a)}}
\int_{x_E}^\infty dx_b\,
\exp\left\{x_b\!-
 \frac{[x_b-x_a-
r_{x_W}(t_b-t_a)]^2}
{2 \sigma^2(t_b-t_a)}
\right\}\! .
\label{@CSprize}\end{eqnarray}
and
\begin{eqnarray}
&&\!\!\!\!\!\!\!\!\!\!\!\!O_E(x_a,t_a)=
Ee^{-r_W(t_b-t_a)}\frac{1}{ \sqrt{ 2\pi \sigma^2(t_b-t_a)}}
\int_{x_E}^\infty dx_b\,
\exp\left\{-
 \frac{[x_b-x_a-
r_{x_W}(t_b-t_a)]^2}
{2 \sigma^2(t_b-t_a)}
\right\} .
\nonumber \\&&
\label{@}\end{eqnarray}
In the second integral we set
\begin{equation}
x_-\equiv
x_a+r_{x_W}(t_b-t_a)=
x_a+\left(r_W-\frac{1}{2} \sigma^2\right)(t_b-t_a),
\label{@}\end{equation}
and obtain
\begin{eqnarray}
\!\!\!\!\!\!\!\!\!\!\!\!\!O_E(x_a,t_a)=
E\frac{e^{-r_W(t_b-t_a)}}{ \sqrt{ 2\pi \sigma^2(t_b-t_a)}}
\int_{x_E-x_-}^\infty dx_b\,
\exp\left\{-
 \frac{x_b^2}
{2 \sigma^2(t_b-t_a)}
\right\} .
\label{@}\end{eqnarray}
After rescaling the integration variable
$x_b\rightarrow -\xi \sigma \sqrt{t_b-t_a} $,
this can be rewritten as
\begin{eqnarray}
O_E(x_a,t_a)=
e^{-r_W(t_b-t_a)}E\,N(y_-),
\label{@}\end{eqnarray}
where
$N(y)$ is the Gaussian distribution function
\begin{equation}
N(y)\equiv \int _{-\infty}^y\frac{d\xi}{ \sqrt{2\pi} }
e^{-\xi^2/2}.
\label{@}\end{equation}
evaluated at
\begin{eqnarray}
y_{-}&\equiv&\frac{ x_--x_E}{ \sqrt{ \sigma^2(t_a-t_b)} }
=\frac{\log[S(t_a)/E]+r_{x_W}(t_b-t_a)}{ \sqrt{ \sigma^2(t_a-t_b)} }
\nonumber \\&=&\frac{\log[S(t_a)/E]+\left(r_W-\frac{1}{2} \sigma^2\right)(t_b-t_a)}{ \sqrt{ \sigma^2(t_a-t_b)} }.
\label{@}\end{eqnarray}

The integral in the
first contribution
(\ref{@CSprize}) to
    the option price
is found after
completing
the exponent
in the integrand
quadratically as follows:
\begin{eqnarray}
&&\!\!\!\!\!\!\,\,\!\!\!\!\!\!\!\!\!\!\!\!\!\!\!\!\!\!\!\!\!\!\!\!x_b-\frac{[x_b-x_a-
r_{x_W}(t_b-t_a)]^2}
{2 \sigma^2(t_b-t_a)}  \nonumber \\&&
\!\!\!\!\!\!\!\!\!\,\,\!\!\!\!\!\!\!\!\!\!\!\!\!\!\!\!\!\!\!~~~=
-\frac{[x_b-x_a-(r_{x_W}+ \sigma^2)(t_b-t_a)
]^2-2r_W \sigma^2(t_b-t_a)-2x_a \sigma^2(t_b-t_a)}
{2 \sigma^2(t_b-t_a)}.
\label{@}\end{eqnarray}
Introducing now
\begin{equation}
x_+\equiv
x_a+\left(r_{x_W}+ \sigma^2\right)(t_b-t_a)=
x_a+\left(r_W+\frac{1}{2} \sigma^2\right)(t_b-t_a),
\label{@}\end{equation}
and rescaling $x_b$ as before, we obtain
\begin{equation}
O_S(x_a,t_a)=S(t_a)N(y_+),
\label{@}\end{equation}
%
%
with
\begin{eqnarray}
y_{+}&\equiv&
\frac{x_+-x_E}{ \sqrt{ \sigma^2(t_a-t_b)} }
=\frac{\log[S(t_a)/E]+\left(r_{x_W}+ \sigma^2\right)(t_b-t_a)}{ \sqrt{ \sigma^2(t_a-t_b)} }
\nonumber \\&=&\frac{\log[S(t_a)/E]+\left(r_W+\frac{1}{2} \sigma^2\right)(t_b-t_a)}{ \sqrt{ \sigma^2(t_a-t_b)} }.
\label{@}\end{eqnarray}
The combined result
\begin{eqnarray}
O(x_a,t_a)=S(t_a)N(y_+)-
e^{-r_W(t_b-t_a)}E\,N(y_-)
\label{@BLACKSCH}\end{eqnarray}
is the celebrated
{\em Black-Scholes formula}%
\iems{Black-Scholes+formula}\iems{formula,Black-Scholes}
of option pricing.

\begin{figure}[bth]
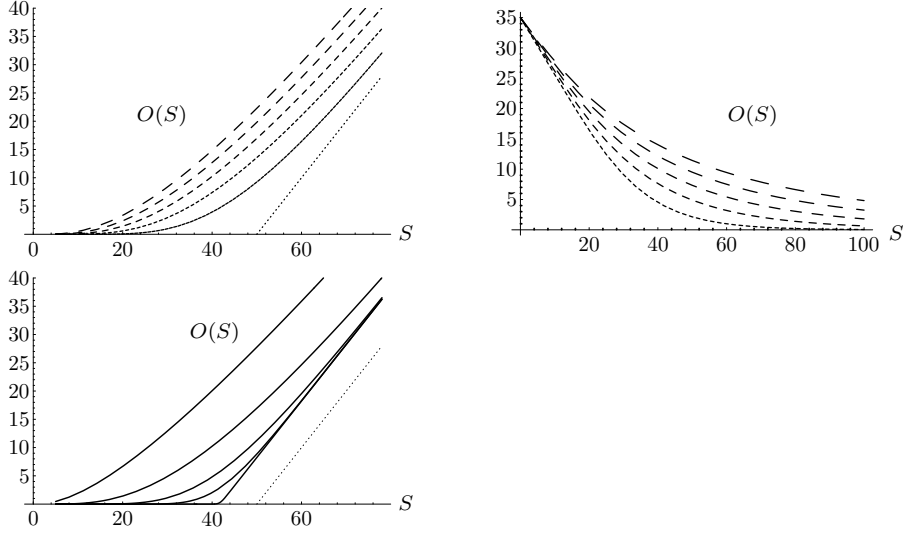

~\\[2cm] \phantom{x}\input pls.tps \hspace{5cm}
\input strike.tps 
\hspace{-.11cm}
~\\[1.cm]
 ~~~\phantom{x}\input plsi.tps
~\\[-1.cm]
\caption[Dependence of call price $O$ on the stock price $S$,
the strike price $E$, and the volatility
$ \sigma $]
{Left: Dependence of call price $O$ on the stock price $S$
for different times before expiration date
(increasing dash length:
$1,2,3,4,5$ months).
from left to right:
$1,2,3,5,6$ months).
The parameters are $E=50$\,US\$, $ \sigma=40\%$, $r_W=6\%$ per month.
Right: Dependence on the strike price $E$
for fixed stock price  $35$\,US\$ and the
same times to expiration
(increasing with dash length).
Bottom: Dependence on the
 volatilities (from left to right:
$80\%,60\%,20\%,10\%,1\%$) at a fixed time $t_b-t_a=3 $ months
before expiration.
}
\label{@deponS}\end{figure}

In Fig.~\ref{@deponS}
we illustrate how the dependence of the call price
on the stock price
varies
with different times to expiration $t_b-t_a$ and
with different volatilities $ \sigma$.
~\\

\subsection{Option Pricing for Non-Gaussian Fluctuations}

For non-Gaussian
fluctuations,
the option price must be calculated numerically
from Eqs.~(\ref{@BSINTEGRAL}) and (\ref{@strikepr}).
Inserting the Fourier representation
(\ref{@18stprgf}) and using the
Hamiltonian
\begin{equation}
H_{r_{x_W}}(p)\equiv \bar H(p)+ir_{x_W}p
\label{@HLevyB}\end{equation}
defined as in
 (\ref{@cumuatsHar}), this becomes
\begin{eqnarray}
\!\!\!\!\!\!\!\!\!\!\!\!\!\!\!\!O(x_a,t_a)&=&
\int_{x_E}^\infty  dx_b\,(e^{x_b}-e^{x_E})
 P(x_bt_b|x_at_a)\nonumber \\
&=&
e^{-r_W(t_b-t_a)}\int_{x_E}^\infty  dx_b\,(e^{x_b}-e^{x_E})
\int _{-\infty }^\infty
 \frac{dp}{2\pi}e^{ip(x_b-x_a)- H_{r_{x_W}}(p)
 (t_b-t_a)} .  ~~~~
\label{@opprtLevyB}\end{eqnarray}
The integrand can be rearranged as follows:
\begin{equation}
O(x_a,t_a)\!=\!e^{-r_W(t_b-t_a)}  \!
\int_{x_E}^\infty \! dx_b\int_{-\infty}^\infty
 \frac{dp}{2\pi}\left[
e^{x_a}e^{(ip+1)(x_b-x_a)}
\!-e^{x_E}e^{ip(x_b-x_a)}\right]e^{- H_{r_{x_W}}(p)
 (t_b-t_a)} ,
\label{@18stA}\end{equation}
Two integrations are required.
This would make a numerical calculation quite time consuming,
Fortunately, one integration can be done
analytically.
For this purpose
we
write the integral in the form
\begin{equation}
O(x_a,t_a)=e^{-r_W(t_b-t_a)}
\int_{x_E}^\infty  dx_b\int_{-\infty}^\infty
 \frac{dp}{2\pi}
e^{ip(x_b-x_a)}
f(x_a,x_E;p),
\label{@18stB1}\end{equation}
with
\begin{equation}
\!\!\!\!\!\!\!\!\!f(p)\equiv
e^{x_a}e^{- H_{r_{x_W}}(p+i)
 (t_b-t_a)}
-e^{x_E}e^{- H_{r_{x_W}}(p)
 (t_b-t_a)}
 .
\label{@18stB}\end{equation}
We have suppressed the arguments
$x_a,x_E,t_b-t_a$ in $f(p)$, for brevity.
The integral over $x_b$  in (\ref{@18stB1})
runs over the Fourier transform
\begin{equation}
 \tilde f(x_b-x_a)=
\int_{-\infty}^\infty
 \frac{dp}{2\pi}
e^{ip(x_b-x_a)}
f(p),
\label{@20FR}\end{equation}
of the function $f(p)$.
It is then convenient
to express the integral $\int_{x_E}^\infty  dx_b$
in terms of the Heaviside function
$\Theta(x_b-x_E)$ as
$\int_{-\infty }^\infty  dx_b \, \Theta(x_b-x_E)$ and use
the Fourier representation
\begin{equation}
\Theta(x_b-x_E)=
\int \frac{dq}{2\pi}\frac{i}{q+i \eta }e^{-iq(x_b-x_E)}.
\label{@}\end{equation}
 of the Heaviside function
to write
\begin{equation}
 \int_{x_E}^\infty  dx_b\, \tilde f(x_b-x_a)= \int_{-\infty }^\infty  dx_b
\int \frac{dq}{2\pi}\frac{i}{q+i \eta }e^{-iq(x_b-x_E)}\tilde f(x_b-x_a).
\label{@}\end{equation}
Inserting here
the Fourier representation
(\ref{@20FR}), we can perform the integral over $x_b$
and obtain the momentum space representation
of the option price
\begin{eqnarray}
O(x_a,t_a)=
\int_{-\infty}^\infty
 \frac{dp}{2\pi}
e^{ip(x_E-x_a)}
\frac{i}{p+i \eta }
{f(p)}.
\label{@20DRoptpr}\end{eqnarray}
For numerical integrations, the singularity at $p=0$
is inconvenient.
We therefore use the well-known decomposition
\begin{equation}
\frac{i}{p+i \eta }=\frac{{\cal P}}p+ \pi  \delta (p),
\label{@}\end{equation}
to write
\begin{eqnarray}
O(x_a,t_a)=  \frac{1}{2} f(0)+i
\int_{-\infty}^\infty
 \frac{dp}{2\pi}
\frac{{
e^{ip(x_E-x_a)}
f(p)}-f(0)}p.
\label{@20DRoptpr1}\end{eqnarray}
We have used the fact that the principal value of the integral
over $1/p$ vanishes to subtract
the constant $f(0)$ from $
e^{ip(x_E-x_a)}
 f(p)$.
After this the
 integrand
is
regular, does not need any more the principal-value
specification,  and allows for a numerical integration.

For $x_a$ very much different from $x_E$,
we may approximate
\begin{equation}
\int_{-\infty}^\infty
 \frac{dp}{2\pi}
\frac{{
e^{ip(x_E-x_a)}
f(p)}-f(0)}p\approx \frac{1}{2}\epsilon(x_a-x_E)f(0),
\label{@}\end{equation}
where $ \epsilon (x)\equiv 1+2\Theta(x)$ is the step function,
and obtain
\begin{eqnarray}
O(x_a,t_a)\approx
\frac{1}{2}\left[1+\Theta(x_a-x_E)\right]f(0)
.
\label{@20DRoptpr2}\end{eqnarray}
Using (\ref{@ratexv}) we have
$e^{- H_{r_{x_W}}(i)}=
e^{r_W(t_b-t_a)}$, and since
$e^{- H_{r_{x_W}}(0)}= 1$,
we see that $O(x_a,t_a)$ goes to zero for
$x_a\rightarrow -\infty$ and
has the large-$x_a$
 behavior
\begin{eqnarray}
O(x_a,t_a)\approx
\left(e^{x_a}-e^{x_E}
e^{-r_W(t_b-t_a)}
\right)= S(t_a)-e^{-r_W(t_b-t_a)}E
.\label{@20DRoptpr2}\end{eqnarray}
This is the same behavior as in the Black-Scholes formula
(\ref{@BLACKSCH}).

\begin{figure}[b]
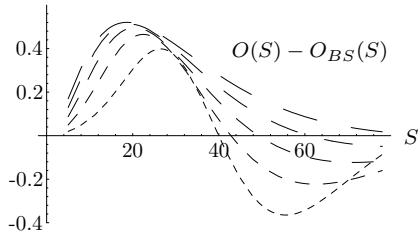

~\\[2.5cm] \phantom{x}\input plsitld.tps
\hspace{3cm}
\vspace{-.3cm}
\caption[Dependence of call price $O$ on the stock price $S$
for \tld]
{Difference of the call price $O$
obtained from \tld{} with kurtosis $ \kappa =4$ and the
Black-Scholes price as a function of the stock price $S$
for different times before expiration date (increasing dash length:
$1,2,3,4,5$ months).
The parameters are $E=50$\,US\$, $ \sigma=40\%$, $r_W=6\%$ per month.
}
\label{@deponSD}\end{figure}

In Fig.~\ref{@deponSD}
we display the difference between the option prices
emerging from   our
formula (\ref{@20DRoptpr1}),
with a \tld{} of kurtosis $ \kappa =4$, and the Black-Scholes formula
(\ref{@BLACKSCH}) for the same data
 as in the upper left of Fig.~\ref{@deponS}.

\section[]{Conclusion}

The  stochastic calculus
and the option pricing formulas
developed in this paper
will be useful for
estimating financial risks
of a variety of investments.
In particular,
it will help developing a more realistic theory of
fair option prices.
More details can be found in
the textbook Ref.~\cite{PI}.

~\\           ~\\
~\\           ~\\
Acknowledgment

The author is grateful to Marc Potters,
Ernst Eberlein, Jan Kallsen, Michail Nechaev,
Peter Bank,
and
Matthias Otto
 for useful discussions.
He also  thanks Axel Pelster and  Flavio Nogueira
for  useful comments.

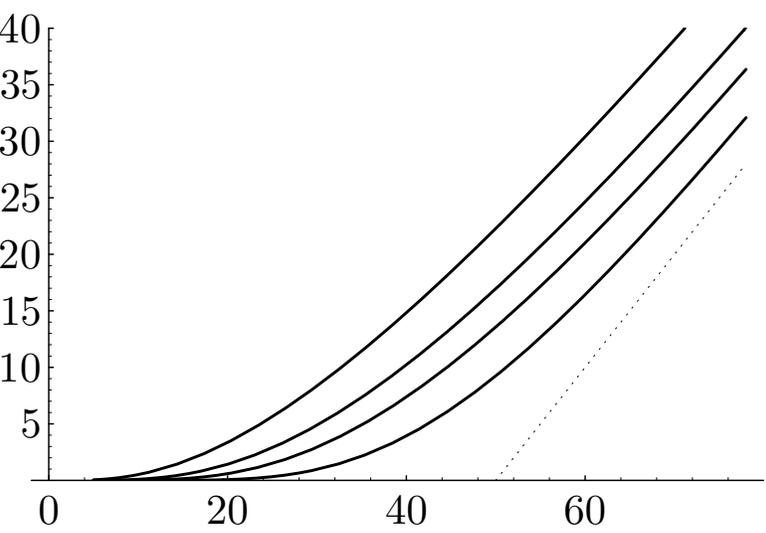
\end{document}